\documentclass[a4paper,12pt]{article} 
\usepackage{epsfig}
\usepackage{amsmath,amsthm, amssymb, amsfonts}
\usepackage{graphicx}
\usepackage{color}

\newtheorem*{prof*}{Proof}

\newcommand{\btb}{\begin{table}}
\newcommand{\etb}{\end{table}}
\newcommand{\bt}{\begin{Theorem}}
\newcommand{\et}{\end{Theorem}}
\newcommand{\bp}{\begin{Proof}}
\newcommand{\ep}{\end{Proof}}
\newcommand{\bl}{\begin{Lemma}}
\newcommand{\el}{\end{Lemma}}
\newcommand{\br}{\begin{Remark}}
\newcommand{\er}{\end{Remark}}
\newcommand{\bi}{\begin{itemize}}
\newcommand{\ei}{\end{itemize}}
\newcommand{\bean}{\begin{eqnarray*}}
\newcommand{\eean}{\end{eqnarray*}}
\newcommand{\be}{\begin{equation}}
\newcommand{\ee}{\end{equation}}
\newcommand{\ben}{\begin{equation*}}
\newcommand{\een}{\end{equation*}}

\newcommand{\bc}{\begin{center}}
\newcommand{\ec}{\end{center}}

\begin{document}
\title{\textbf{Understanding Betting Strategy}}

\author{K\lowercase{anika} S\lowercase{aha}$^{1}$ \lowercase{and} A\lowercase{nanya} L\lowercase{ahiri}$^{\dag,2}$\footnote{\dag{\it Corresponding author}, {E-mail: \tt ananya.isi@gmail.com}}
\\
$^{1}$ A\lowercase{lgo} L\lowercase{ab} CMI, I\lowercase{ndia}\\
$^{2}$C\lowercase{hennai} M\lowercase{athematical} I\lowercase{nstitute}, C\lowercase{hennai}, I\lowercase{ndia}}


\date{05 June 2017}

\maketitle

\begin{abstract}
In this paper, we present betting strategy of a football game using probability theory. We know all betting houses offer slightly unfair odds towards the player. Here we discuss a simple way to figure out which betting house is offering relatively better odds compared to others for English Premier League. However, this methodology can be adopted for another league football match.
\end{abstract}

\section{Introduction}

It is believed that betting house always makes money in long run irrespective of their short term loss or gain.  In this paper, we make an attempt to understand this phenomenon with the concept of simple `expectation' and `variance' of probability theory. First, we will discuss what is \textbf{fair game}. Let's consider a simple game of English Premier League (EPL) where Manchester United (ManU) is playing against Liverpool. Suppose a betting house offers a game that if ManU wins with probability 0.606, then the player will receive \$0.65 from the betting house. On the other hand, if ManU loses with probability 0.394 then the player has to pay \$1 to the betting house.

Now player's revenue scheme $R_p$ is defined as follows:
\begin{eqnarray*}
R_{p}=\bigg\{\begin{array}{cc}
              0.65 & \mbox{with probability }0.606,\\
              -1   & \mbox{with probability }0.394,
            \end{array}
\end{eqnarray*}
and player's expected revenue is 
$$
E(R_p)=0.65\times 0.606 - 0.394=0.
$$
More about expectation and moments can be found in [1,2]. Now, we will look at the revenue scheme and expected revenue for the betting house for the same game.
\begin{eqnarray*}
R_{b}=\bigg\{\begin{array}{cc}
              -0.65 & \mbox{with probability }0.606,\\
              1   & \mbox{with probability }0.394.
            \end{array}
\end{eqnarray*}
Expected revenue for betting house is,
$$
E(R_b)=-0.65\times 0.606 + 0.394=0.
$$
If we compare $R_p$ and $R_b$, then we can see player's loss is the gain of the betting house and vice-versa.
This type of game is known as `zero-sum' game. Also, $E(R_b)=E(R_p)=0$ means if the player and betting house play this game several times, then `on-average' neither betting house nor the player will make or lose money. This is called `fair-game.'

Now we know that the betting house has an establishment cost and most of the time these betting houses are the for-profit organization. Here the question is how they are making money. An easy way of doing it, if the betting house pays the player less than what they suppose to pay. That means, in the fair game, if they think to pay the player \$0.65, but to make money, they will pay the player less than \$0.65. For example, if they pay the player \$0.6 then the revenue scheme for the betting house is,
\begin{eqnarray*}
R_{b}=\bigg\{\begin{array}{cc}
              -0.6 & \mbox{with probability }0.606,\\
                1   & \mbox{with probability }0.394,
            \end{array}
\end{eqnarray*}
and the expected revenue for betting house is,
$$
E(R_b)=-0.6\times 0.606 + 0.394=0.0304.
$$
It means if the player and betting house play this game several times, the on-average betting house will make \$0.03 or 3 cents from the player and the player will lose the same amount because it is a zero-sum game. So to earn money, in the long run, the betting house offers the player less than what is fair. Rest of the paper is organized as follows. In section \ref{sec_betting_strategy_prob_th}, we present the strategy of betting houses using probability theory. In section \ref{sec_strategy_league_football_match}, we discuss the strategy for league football match. In section \ref{sec_data_analysis}, we presented the data analysis and showed how imputed cost of a betting house could be estimated numerically.

\section{Betting Strategy with Probability Theory}\label{sec_betting_strategy_prob_th}

In this section, we present the strategy of the betting house. Suppose $A$ is an event with $P(A)=p$. If $A$ happens then, the betting house will pay \$$r$. Otherwise, the betting house will receive \$1. So the revenue scheme for the betting house is 
$$
R_{b}=\bigg\{
            \begin{array}{cc}
              -r & \mbox{with probability }p,\\
                1   & \mbox{with probability }(1-p),
            \end{array}
$$
and the expected revenue for betting house is,
$$
E(R_b)=-rp+(1-p).
$$
Now this game is fair game if $E(R_b)=0$, that is $r=\frac{1}{p}-1$,
and $\frac{1}{p}$ is known as `\emph{decimal odds}'. The variance of the revenue scheme is generally considered as risk of a game. So for a fair game $Var(R_b)=E(R_b^2)-[E(R_b)]^2=E(R_b^2)$. Now

\begin{eqnarray*}
E(R_b^2)&=& p \Big(\frac{1}{p}-1\Big)^2+ (1-p)\\
&=& (1-p)\Big[\frac{1-p}{p}+1\Big]\\
&=&\frac{1-p}{p}=\frac{1}{p}-1=r.
\end{eqnarray*}

Interestingly, for the fair game, $r$ is the amount which betting house pays for each dollar they receive. It turns out that $r$ is also the measure of risk for the same strategy as $Var(R_b)=r$, which is known as `\emph{fractional odds}.' To make a profit, in the long run, betting house pays \$$(r-\epsilon)$, where $\epsilon>0$. The revenue scheme is 
$$
R_{b}=\bigg\{
            \begin{array}{cc}
              -\{(\frac{1}{p}-1)-\epsilon\} & \mbox{with probability }p,\\
                1   & \mbox{with probability }(1-p),
            \end{array}
$$
and the expected revenue of betting house is,
\begin{eqnarray*}
E(R_b)&=&-(\frac{1}{p}-1)p+\epsilon p+ (1-p)\\
&=&-(1-p)+\epsilon p + (1-p)\\
&=&\epsilon p.
\end{eqnarray*}
After simplification $Var(R_{b})=E(R_b^2)-[E(R_b)]^2= r(1-p\epsilon)^2$.

\section{Strategy for League Football Match}\label{sec_strategy_league_football_match}
There are three mutually exclusive outcomes in a league football match. The outcomes are (i) Hometeam win, (ii) Away team win, (iii) Draw. Let us consider the EPL match between ManU Vs. 
Liverpool, where ManU is Hometeam.  Suppose for this match betting house calculates the probabilities 0.6, 0.15 and 0.25 for three outcomes, namely Hometeam wins, Away team wins and draw respectively. The corresponding decimal odds are $1/0.6=1.66$, $1/0.15=6.66$ and $1/0.25=4.00$.

In the previous section, we noticed that the betting house would never reveal these fair odds. They will always announce odds which are less than the fair value so that they can stay in profit. For instance, if the betting house offers odds of 1.57, 6.57 and 3.87 against the respective events of Hometeam wins, Away team wins and draw, then the revised revenue scheme for ManU to win the match with announced odds of 1.57 is,
$$
R_{b}^H=\bigg\{
            \begin{array}{cc}
              -(1.57-1) & \mbox{with probability }0.6,\\
                1   & \mbox{with probability }0.4.
            \end{array}
$$
The expected revenue for the betting house is,
$$
E(R_b^H)=-0.57\times 0.6 +  0.4=0.058.
$$
Similarly, the revised revenue scheme for the Liverpool to win the match with announce odds of 6.57 is,
$$
R_{b}^A=\bigg\{
            \begin{array}{cc}
              -(6.57-1) & \mbox{with probability }0.15,\\
                1   & \mbox{with probability }0.85,
            \end{array}
$$
and the expected revenue for betting house is,
$$
E(R_b^A)= -5.57\times 0.15 + 0.85=0.0145.
$$
Likewise, the revised revenue scheme that match will be draw with announce odds of 3.87 is,
$$
R_{b}^D=\bigg\{
            \begin{array}{cc}
              -(3.87-1) & \mbox{with probability }0.25,\\
                1   & \mbox{with probability }0.75,
            \end{array}
$$
and the expected revenue for betting house is,
$$
E(R_b^D)= -2.87\times 0.25 + 0.75=.0325.
$$

Expected revenue of betting house for all events are positive. Mathematically, we can show when the announce odds are converted to probabilities; they usually add up to more than 1 to keep the house on benefit.

Let us assume, $\Big(\frac{1}{P_H}-\epsilon\Big)=\frac{1}{P_{H}^*},$ $\Big(\frac{1}{P_A}-\epsilon\Big)=\frac{1}{P_{A}^*}$ and $\Big(\frac{1}{P_D}-\epsilon\Big)=\frac{1}{P_{D}^*}$ 

Therefore,
\begin{eqnarray*}
\frac{1-\epsilon P_H}{P_H}=\frac{1}{P_H^*}\\
\frac{P_H}{1-\epsilon P_H}={P_H^*}\\
{P_H}<{P_H^*}
\end{eqnarray*}

In the same way we can show ${P_A}<{P_A^*}$ and ${P_D}<{P_D^*}$. Now adding both sides we can conclude,

${P_H}+{P_A}+{P_D}<{P_H^*}+{P_A^*}+{P_D^*}$.

Of course probabilities of a fair game sums up to 1 i.e. ${P_H}+{P_A}+{P_D}=1$, hence evidently ${P_H^*}+{P_A^*}+{P_D^*}>1$.

\section{Data Analysis}\label{sec_data_analysis}

Let us discuss the former analysis above with EPL data. Data is accessible on \textbf{http://www.football-data.co.uk/englandm.php}. In the table (\ref{table_B365}) we present the decimal odds of 10 different EPL matches from the popular betting house Bet365. First three columns, `B365H', `B365D' and `B365A' represent the decimal odds (i.e. $1/p$) of three events that are (i) home team wins, (ii) draw and (iii) away team wins; from the betting house Bet365. It is visible that total probability of the three mutually exclusive events of a given match is greater than one which violates the thumb rule that probabilities add up to one. It is inevitable that sum of the probabilities will be greater than one from our results described in the previous section. However, if there are two betting houses, we can say that one house is offering better odds to the players if its sum of probabilities is closer to one.


Let us consider a match held on 8th August 2015, between Bournemouth and Aston Villa where the former team is Hometeam and later one is Away team.  We look at the betting houses for this match, the following table \ref{table_compare_odds} (for the six betting houses, namely Bet365 (B365), Bet\&Win (BW), Interwetten (IW), Ladbrokes (LB), William Hill (WH), VC Bet (VC) ). Here we can compare the announce odds whichever is near to 1 that betting the house is offering fair. As per the table \ref{table_compare_odds}, B365 is offering fair odds compared with others. The betting house Interwetten is offering worst odds for this match. 

\subsection{Imputed Cost of a Betting House}

We can show different ways of calculating Imputed cost of the betting house. Here we are using additive model for the same. In this model, revenue of betting house is described as,
$$
R_{b}=\bigg\{
            \begin{array}{cc}
              -\{(\frac{1}{p}-1)-\epsilon\} & \mbox{with probability }p,\\
                1   & \mbox{with probability }(1-p).
            \end{array}
$$
As discussed earlier, to make a profit, betting houses will always offer an amount which is less than what they suppose to pay. Here $\epsilon$ is that profitable amount. Suppose the betting houses announce odds are ${P_{H}^*}$, ${P_{A}^*}$ and ${P_{D}^*}$ respectively for Hometeam, Away team and Draw of a match. These announce odds can be defined as below:
$$\Big(\frac{1}{P_H}-\epsilon\Big)=\frac{1}{P_{H}^*},~~
\Big(\frac{1}{P_A}-\epsilon\Big)=\frac{1}{P_{A}^*},~~
\Big(\frac{1}{P_D}-\epsilon\Big)=\frac{1}{P_{D}^*}$$
Now we all know that probabilities of three mutually exclusive events add up to 1 for fair game, i.e. $P_{H}+P_{A}+P_{D}=1$. But ${P_{H}^*}+{P_{A}^*}+{P_{D}^*}>1$. Let us say $\delta$ is the difference between sum of these two probabilities. Then

\begin{eqnarray*}
&&{P_H^*}+{P_A^*}+{P_D^*}-({P_H}+{P_A}+{P_D})=\delta\\
&\Rightarrow
&{P_H^*}+{P_A^*}+{P_D^*}-(\frac {P_H^*}{1+\epsilon P_H^*}+\frac {P_A^*}{1+\epsilon P_A^*}+ \frac {P_D^*}{1+\epsilon P_D^*})=\delta\\
&\Rightarrow
&{P_H^*}-\frac {P_H^*}{1+\epsilon P_H^*}+{P_A^*}-\frac {P_A^*}{1+\epsilon P_A^*}+{P_D^*}-\frac {P_D^*}{1+\epsilon P_D^*}=\delta\\
&\Rightarrow
&{P_H^*}(1-\frac{1}{1+\epsilon{P_H^*}})+{P_A^*}(1-\frac{1}{1+\epsilon{P_A^*}})+{P_D^*}(1-\frac{1}{1+\epsilon {P_D^*}})=\delta\\
&\Rightarrow
&\frac {\epsilon (P_H^*)^2}{1+\epsilon P_H^*} + \frac {\epsilon (P_A^*)^2}{1+\epsilon P_A^*} + \frac {\epsilon (P_D^*)^2}{1+\epsilon P_D^*}= \delta.
\end{eqnarray*}

Clearly we can estimate $\epsilon$ from the above equation as ${P_H^*},{P_A^*},{P_D^*}$ and $\delta$ are already known to us. Let us solve the equation for $\epsilon$ using the given EPL data. Considering the first match for B365 betting house where $\{P_H^*\}=0.50$, $\{P_A^*\}$=0.25, $\{P_D^*\}$=0.28 and $\delta$=1.03-1=0.03. Now solving the above equation with the help of this numerical values,
\begin{eqnarray*}
&\epsilon(\frac {0.50^2}{1+\epsilon (0.50)} + \frac {0.25^2}{1+\epsilon (0.25)} + \frac {0.28^2}{1+\epsilon (0.28)})=0.03.
\end{eqnarray*}
Here we can see the equation is cubic in $\epsilon$. Instead solving analytically which is complicated, we are introducing \texttt{R} programming to optimise $\epsilon$.
Therefore calculate the value of $\epsilon$ is 0.065 or following the revenue model we can say \$0.065 per dollar is the profitable amount for the betting house.

\subsection{Comparative study of Imputed cost among different betting houses: }

\begin{figure}
\label{fig_imputed_cost}
\centering
\caption{Imputed Cost of 380 matches by Different betting Houses}
\includegraphics[height=10in,width=7in]{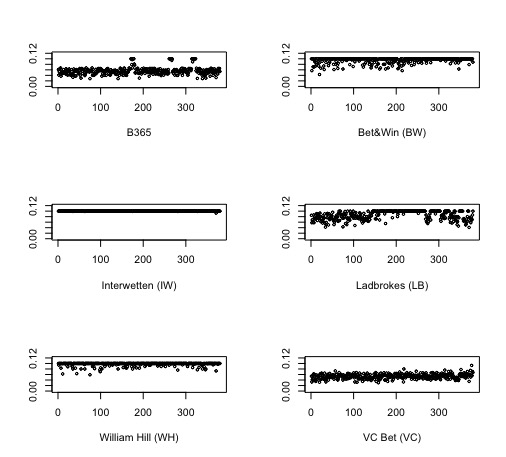}
\end{figure}

Let us consider the result presented in Table (\ref{table_imputed_cost}) and figure (\ref{fig_imputed_cost}). 
\begin{enumerate}
\item All betting houses are limited to \$0.10 seems like there is some regulatory policy
\item For $95\%$ of the matches B365 charges between \$.04 to \$.06.
\item For majority games and approximately for all matches BW and IW respectively charge \$0.10.
\item Same behavior we can see for LB also.
\item But VC never charges greater than \$.09. Moreover, for majority matches, they charge between \$.04 to \$.06. 
\item B365 and VC has similar kind of behavior.
\end{enumerate}

\section{Conclusion}

In this paper, we presented how the strategy of the betting houses can be explained with the probability odds. From the real data, it is visible that for all betting houses considered here, the sum of the probability for the win, loss, and draw of a match are greater than one. We also presented how the cost of a game can be imputed numerically. Therefore, with this understanding, one realizes how the betting houses are consistently making money and accordingly decides whether to invest one's money on the betting house.


\begin{thebibliography}{9}
\bibitem{FellerVol1} 
Feller, W.
\textit{An Introduction to Probability Theory and Its Applications}.
Wiley Series in Probability and Statistics, Volume 1, 1968.
 
\bibitem{CRRAO} 
Rao, C. R, \textit{Linear Statistical Inference and its Applications}. 
Wiley Series in Probability and Statistics, 2nd Edition ,ISBN: 978-0-471-21875-3, 2001.

\end{thebibliography}

\section*{Appendix: Figures and Tables}

\begin{table}[ht]
\caption{Table here}\label{table_B365}
\centering
\begin{tabular}{rrrrrrrr}
  \hline
 & B365H & B365A & B365D & P(B365H) & P(B365A) & P(B365D) & Sum of prob \\ 
  \hline
1 & 2.00 & 4.00 & 3.60 & 0.50 & 0.25 & 0.28 & 1.03 \\ 
  2 & 1.36 & 11.00 & 5.00 & 0.74 & 0.09 & 0.20 & 1.03 \\ 
  3 & 1.70 & 5.50 & 3.90 & 0.59 & 0.18 & 0.26 & 1.03 \\ 
  4 & 1.95 & 4.33 & 3.50 & 0.51 & 0.23 & 0.29 & 1.03 \\ 
  5 & 1.65 & 6.00 & 4.00 & 0.61 & 0.17 & 0.25 & 1.02 \\ 
  6 & 2.55 & 3.00 & 3.30 & 0.39 & 0.33 & 0.30 & 1.03 \\ 
  7 & 1.29 & 12.00 & 6.00 & 0.78 & 0.08 & 0.17 & 1.03 \\ 
  8 & 2.88 & 2.70 & 3.30 & 0.35 & 0.37 & 0.30 & 1.02 \\ 
  9 & 3.40 & 2.30 & 3.40 & 0.29 & 0.43 & 0.29 & 1.02 \\ 
  10 & 5.75 & 1.67 & 4.00 & 0.17 & 0.60 & 0.25 & 1.02 \\ 
   \hline
\end{tabular}
\end{table}

\begin{table}[ht]\label{table_compare_odds}
\caption{Compare Betting Odds Across Betting Houses for the Same Match}
\centering
\begin{tabular}{rrrrrrrr}
  \hline
 & Hometeam & Awayteam & Draw & P(Hometeam) & P(Awayteam) & P(Draw) & Sum of prob \\ 
  \hline
B365 & 2.00 & 4.00 & 3.60 & 0.50 & 0.25 & 0.28 & 1.03 \\ 
  BW & 2.00 & 3.70 & 3.30 & 0.50 & 0.27 & 0.30 & 1.07 \\ 
  IW & 2.10 & 3.30 & 3.30 & 0.48 & 0.30 & 0.30 & 1.08 \\ 
  LB & 2.05 & 4.00 & 3.30 & 0.49 & 0.25 & 0.30 & 1.04 \\ 
   \hline
\end{tabular}
\end{table}

\begin{table}[ht]\label{table_imputed_cost}
\caption{Summary of Imputed Cost of Six Betting Houses over the 380 League Matches}
\centering
\begin{tabular}{l|rrrrrr}
\hline
 & B365 & BW & IW & LB & WH & VC \\ 
\hline
Mean & 0.06 & 0.09 & 0.10 & 0.09 & 0.10 & 0.05 \\ 
Median & 0.06 & 0.10 & 0.10 & 0.09 & 0.10 & 0.05 \\ 
Maximum & 0.10 & 0.10 & 0.10 & 0.10 & 0.10 & 0.09 \\ 
Minimum & 0.03 & 0.04 & 0.09 & 0.04 & 0.06 & 0.03 \\ 
Sd & 0.01 & 0.01 & 0.00 & 0.02 & 0.01 & 0.01 \\ 
\hline
\end{tabular}
\end{table}

\end{document}